\documentclass{aa}
\usepackage{graphicx}
\usepackage{latexsym}
\usepackage{amsmath}
\usepackage{amssymb}

\newcommand{\MC}{\multicolumn}
\newcommand{\kms}{km\,s$^{-1}$}
\newcounter{qub}
\begin{document}

\title{High-quality Spectrophotometry of the Planetary Nebula in
the Fornax dSph\thanks{%
Based on observations obtained at the European Southern Observatory,
La Silla, Chile (program 072.A-0087(B)).}
}

\author{%
A.Y.~Kniazev\inst{1,2,3}
\and E.K.~Grebel\inst{4,5}
\and S.A.~Pustilnik\inst{3}
\and A.G.~Pramskij\inst{3}
}

\offprints{A.~Kniazev  \email{akniazev@saao.ac.za}}

\institute{
European Southern Observatory, Karl-Schwarzschild-Strasse 2, 85748 Garching, Germany
\and Current address: South African Astronomical Observatory, Observatory Road, Cape Town, South Africa
\and Special Astrophysical Observatory, Nizhnij Arkhyz, Karachai-Circassia,
369167, Russia
\and Astronomical Institute of the University of Basel, Department of Physics
and Astronomy, Venusstrasse 7, CH-4102 Binningen, Switzerland
\and Astronomisches Rechen-Institut, Zentrum f\"ur Astronomie Heidelberg,
University of Heidelberg, M\"onchhofstr.\ 12--14, D-69120 Heidelberg, Germany
}

\date{Received \hskip 2cm January 2006; Accepted \hskip 2cm  March 2007}

\abstract{We present results of NTT spectroscopy of the one known planetary
nebula (PN) in the dwarf spheroidal galaxy Fornax, a gas-deficient Local Group
galaxy that stopped its star formation activity a few hundred million years
ago.  We detected the [\ion{O}{iii}] $\lambda$4363 line with a signal-to-noise
ratio of $\sim$22.  For the first time we detected the weak [\ion{S}{ii}]
$\lambda\lambda$6717,6731 lines (I(6717+6731) $\approx$ 0.01 I(H$\beta$)),
determined the electron number density (N$_e$(SII) = 750 cm$^{-3}$), and
calculated O, N, Ne, Ar, S, Cl, Fe, He and C abundances.  The abundance analysis
presented here is based on the direct calculation of the electron temperature
$T_{\rm e}$ and yields an oxygen abundance of 12+log(O/H) = 8.28$\pm$0.02.
The analysis of the O, Ne, Ar and S abundances shows that the original ISM
oxygen abundance was 0.27$\pm$0.10 dex lower and that third-dredge-up
self-pollution in oxygen took place.  The blue spectrum shows weak Wolf-Rayet
features, and the progenitor star is classified as a weak emission-line
star. Four of the five PNe in dwarf spheroidal galaxies are now known to 
show WR wind
features.  Overall, the metallicity of the progenitor of the PN fits in well
with stellar spectroscopic abundances derived in previous studies as well as
with the stellar age-metallicity relation of Fornax.

 }

\authorrunning{A.Y.\ Kniazev et al.}

\titlerunning{High-quality Spectroscopy of the PN in Fornax}

\maketitle

\section{Introduction}

Understanding how the elemental abundances of galaxies have changed
over time is an essential issue for understanding galaxy evolution.
Abundance measurements constrain theoretical models, providing
important clues on modes and rates of star formation in galaxies and
on the importance of infall and outflows.  \ion{H}{ii} regions indicate
the present-day ($\sim$10 Myr) gas-phase element abundances, while
planetary nebulae (PNe) reveal the chemical composition of a galaxy at
``intermediate'' ages of a few 100 Myr to a few Gyr
(or even up to $\sim$10 Gyr as for the case discussed below).
These nebular
abundances provide information on elements that are not easily
observed in stellar absorption-line spectra.  Ideally one wishes to
combine information from nebular and stellar abundances, but nebular
emission-line spectra have the added advantage of being observable at
high signal to noise within a fraction of the telescope time required
to obtain similarly deep stellar absorption-line data.  For the Local
Group and other nearby galaxies these abundance data can be combined
with star formation histories derived from color-magnitude diagrams of
resolved stars, thereby yielding deeper insights on galaxy evolution,
in particular on the overall chemical evolution of galaxies as a
function of time.

The Local Group consists of three spiral galaxies and a large number
of dwarf galaxies including irregulars, ellipticals, and spheroidals
(see, e.g., Grebel \cite{Grebel01} for a recent review).  The least massive,
least luminous, most gas-deficient and most numerous dwarf galaxies
are the dwarf spheroidals (dSphs).  Only in two dSph galaxies in the
Local Group have PNe been detected to date: one in Fornax (Danziger et
al.\, \cite{Danz78}) and four in Sagittarius (Zijlstra et al.
\cite{ZW96,Z06}).  Searches for PNe in other dSph galaxies yielded no
detections.  For example, no PNe were found with Sloan Digital Sky
Survey (York et al.\, \cite{york00}, Stoughton et al.\, \cite{Stoughton01})
data for the Sextans, Draco and Leo~I dSph galaxies (Kniazev et al.\,
\cite{Ket05b}).  The few PNe known are valuable in order to compare
nebular abundances of gas-deficient galaxies like dSphs with those of
gas-rich galaxies (e.g., dwarf irregulars).  In fact, in gas-deficient
galaxies without H\,{\sc ii} regions, PNe are the only
means of obtaining nebular oxygen abundances (see, e.g., Richer,
McCall, \& Stasi\'nska \cite{RMS98}).  This is of particular importance when
considering metallicity-luminosity relations of galaxies, since
ideally one would like to use comparable metallicity indicators for
these kinds of studies (see, e.g., the discussion in Grebel,
Gallagher, \& Harbeck \cite{GGH03}).

The goal of our present work is to improve our knowledge of the
element abundances of the lone PN in the Fornax dSph through new
high-quality New Technology Telescope (NTT) spectrophotometry.  Fornax
was discovered by Shapley (\cite{Sh38}) and has a distance of about 140 kpc.
With an absolute V-band magnitude of $-13.1$ it is one of the most
luminous dSphs known (e.g., Table 1 in Grebel et al.\, \cite{GGH03}).
It is also one of only two dSphs known to contain globular clusters.
Several of its metal-poor globular clusters are as old as the oldest
Galactic clusters (Buonanno et al.\, \cite{Bu99}), adding
Fornax to the list of nearby galaxies that share a common epoch of
early star formation (Grebel \& Gallagher \cite{GreGal04}).
Fornax contains a prominent old population, but most of its stars
formed at intermediate ages (e.g., Stetson et al.\, \cite{St98}).
Unlike other dSphs, Fornax continued to form stars until approximately
200 Myr ago (Grebel \& Stetson \cite{GreSte99};
Saviane, Held, \& Bertelli \cite{Saviane00}).  Its younger
population is located in its central regions as seems generally to be
the case in dSphs with extended star formation histories (e.g., Grebel
1999, 2000; Harbeck et al.\, \cite{Harb01}). However, in Fornax the young
population shows a peculiarly asymmetric, almost V-shaped distribution
(Stetson et al.\, \cite{St98}). Moreover, the centroids of the distributions
of Fornax' older populations are shifted with respect to each other.
Recent wide-field studies by
Coleman et al.\, (\cite{Coleman04}, \cite{Coleman05}) revealed two
shell-like structures that may indicate a past merger event in Fornax.
Dinescu et al.\, (\cite{Dinescu04}) suggest that the termination of star
formation in Fornax about 200 Myr ago may have been caused by a crossing
of the Magellanic stream, which stripped Fornax of its star-forming material.

The metallicities of the stars in Fornax have been studied repeatedly
using both photometric techniques and spectroscopy.  The photometric
studies revealed a considerable abundance spread with a relatively
high mean metallicity
(e.g., Grebel \cite{Grebel97}; Saviane et al.\, \cite{Saviane00}).

Several spectroscopic studies of Fornax have been carried out
in recent years.  For instance, Pont et al.\, (\cite{Pont04})  measured
metallicities for 117 red giants and found a surprisingly high mean
metallicity of [Fe/H] = $-0.9$ dex with a substantial metal-poor tail.
They also demonstrate that Fornax has a complex age-metallicity
relation and that it underwent considerable enrichment over the past
few Gyr.  A recent high-resolution study of three red giants in Fornax
was conducted by Tolstoy et al.\, (\cite{Tol03}). This work illustrates that
Fornax follows the trend of high (near-solar) [$\alpha$/Fe] ratios at
low [Fe/H] that was found in other dSphs, setting these stars apart
from the ones in the Galactic halo
(e.g., Shetrone, C\^ot\'e, \& Sargent \cite{SCS01}).
Battaglia et al.\ (\cite{Battaglia06}) obtained metallicities
for 562 stars in Fornax and found a wide range of abundances covering
$-2.8$ to $-0.1$ dex in [Fe/H].  The more metal-poor component peaks
at $-1.7$ dex and the more metal-rich one at $-0.9$ dex.  Also, these
authors find the more metal-rich and younger stars to be more centrally 
concentrated in good agreement with earlier studies cited above.
While a wide range in metallicities is quite common even in dSphs 
with predominantly old populations (e.g., Shetrone, C\^ot\'e, \& 
Sargent \cite{SCS01}), the range in stellar abundances in Fornax
exceeds even that found in other dSphs with extended star formation
histories (e.g., Koch et al.\ \cite{Koch06}, \cite{Koch07}) since Fornax
formed stars over almost the entire Hubble time.  Nonetheless it
remains difficult to assign ages to individual red giants even if
their metallicity is known.

The PN in the Fornax dSph galaxy, discovered by
Danziger et al.\, (\cite{Danz78}), remains the only one known in this
dwarf.  Danziger et al.\ obtained the first spectroscopic observations
of it. Their results and those of others are discussed in Section 4.

The contents of this paper are organized as follows.
Section~\ref{Obs_red} gives the description of our observations and of
the data reduction.  Our results are summarized in
Section~\ref{txt:res} and discussed Section~\ref{txt:disc}.
The conclusions drawn from this study are summarized in
Section~\ref{txt:summ}.

\begin{figure*}
\centering
\includegraphics[width=12cm,angle=-90,clip=]{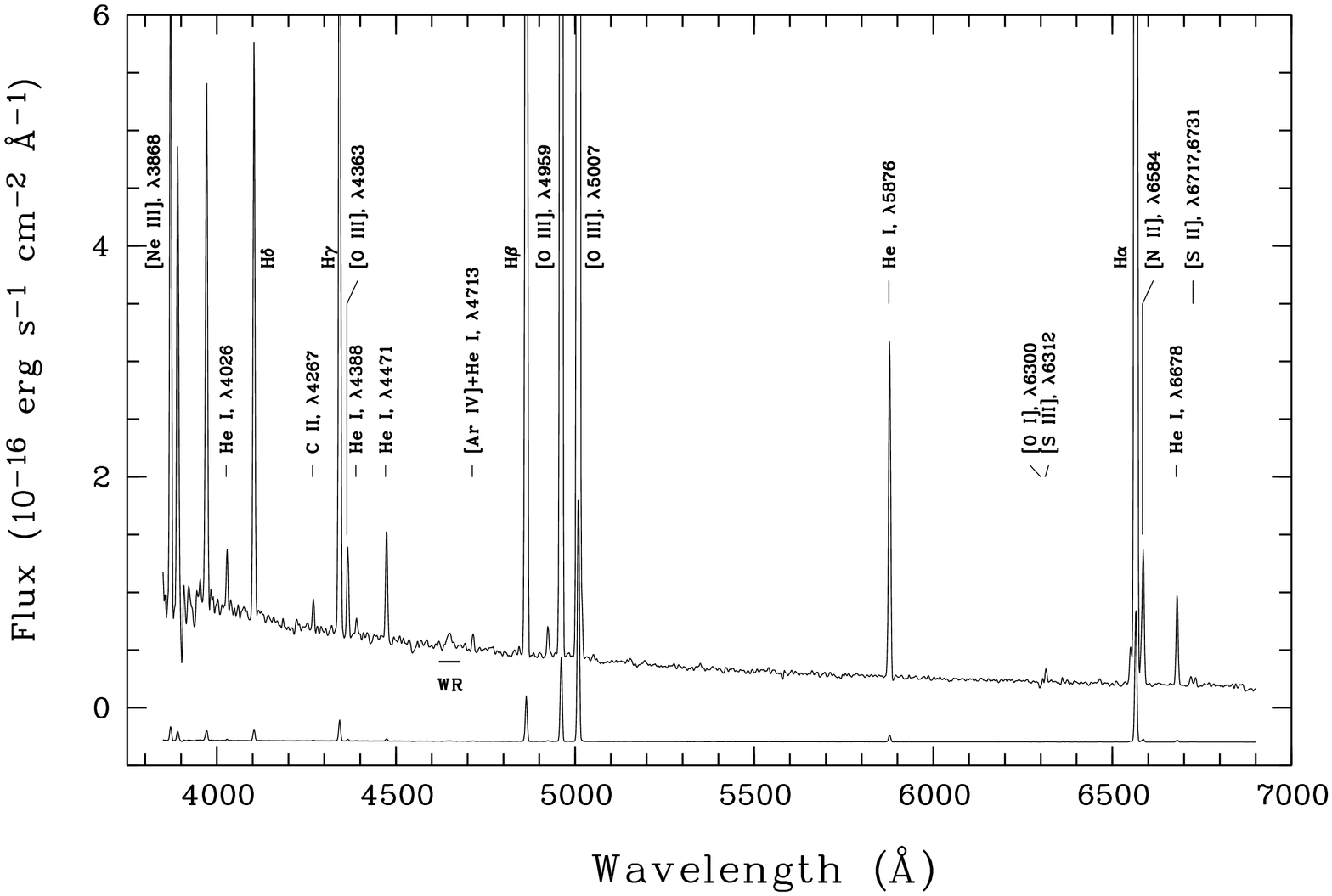}
\includegraphics[width=12cm,angle=-90,clip=]{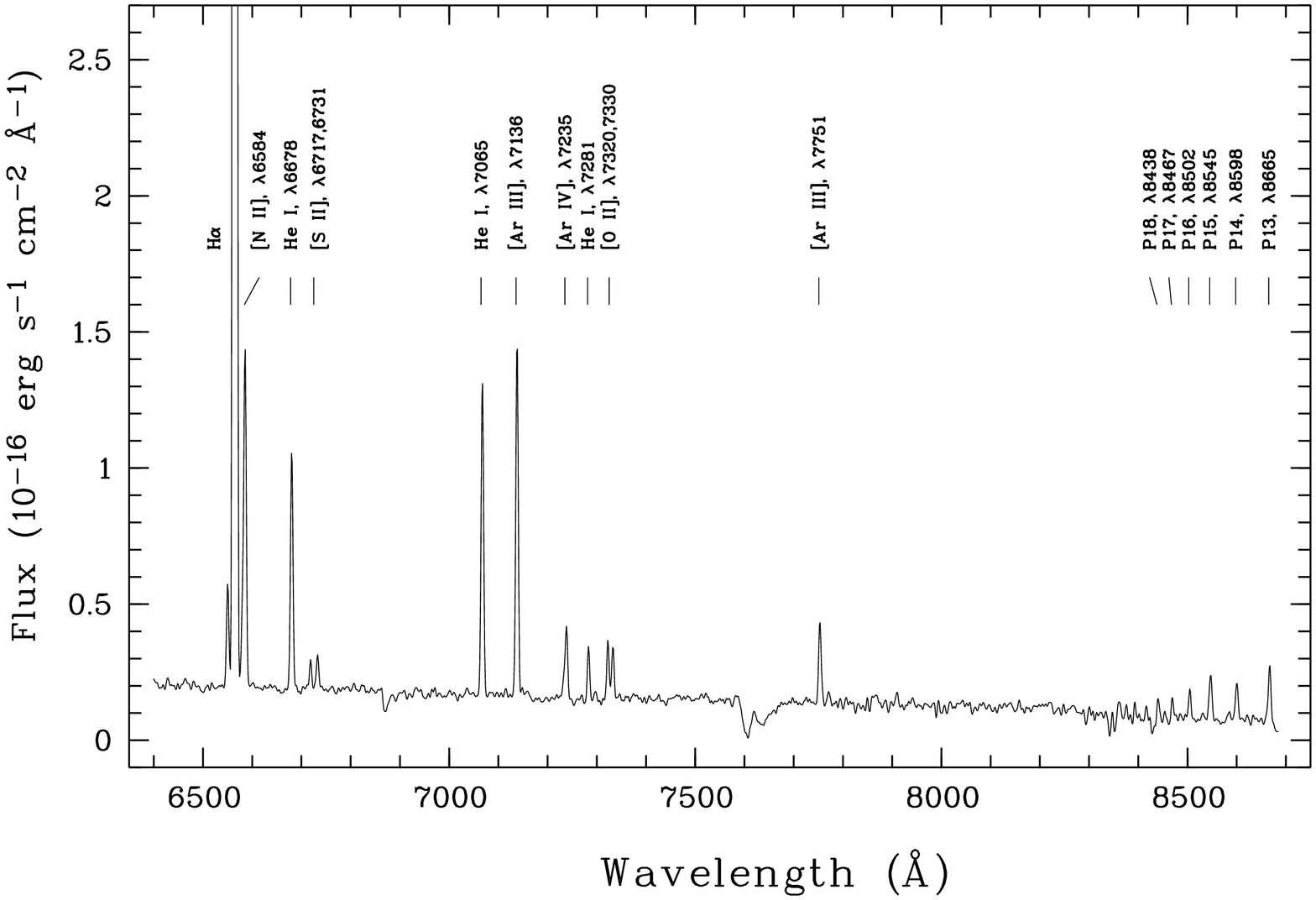}
   \caption{
One-dimensional reduced spectra of the planetary nebula in Fornax that
were obtained with two grisms.  The spectra cover a wavelength range
of 3800--7000 \AA\ (top panel, grism \#5) and 6400--8650 \AA\ (bottom
panel, grism \#6).  Most of the detected emission lines are marked.
The spectrum at the bottom of the top panel is scaled by 1/40 and
shifted to show the relative intensities of the strong lines.  }
\label{fig:PN_spectr} 
\end{figure*}

\section{Spectral Observations and Data Reduction}
\label{Obs_red}

Our spectrophotometric observations were carried out with the NTT at
the European Southern Observatory at La Silla on February 14, 2004.
The observations were performed with the Red Arm of the EMMI
multipurpose instrument using a long slit of 8$^{\prime} \times$
2$^{\prime\prime}$.  For these observations the CCD rows were binned
by a factor of 2, yielding a final spatial sampling of
0.33$^{\prime\prime}$ pixel$^{-1}$.  The seeing during the
observations was very stable and varied within the
range of 0.4$^{\prime\prime}$ to 0.6$^{\prime\prime}$.  The whole
spectral range covered by the two grisms was 3800 -- 8700 \AA\ with a
sampling of 1.63 \AA\ pixel$^{-1}$ in the blue part (3800 -- 7000 \AA;
spectral resolution (FWHM) of $\sim$7 \AA) and with a sampling of
1.4 \AA\ pixel$^{-1}$ in the red part (5750 -- 8670 \AA ; spectral
resolution of $\sim$6.5 \AA).  The PN position in Fornax was taken
from Danziger et al.\, (\cite{Danz78}).  A one-minute H$\alpha$
acquisition image was obtained before the spectroscopic observations
in order to select an optimal position of the slit.  The total
exposure times were 40 minutes for each grism.  The exposures were
broken up into 2 sub-exposures with 20 minutes of exposure time each to
facilitate the removal of cosmic rays.  The airmass changed from 1.3
for the first sub-exposure to 1.75 for the last sub-exposure.  Spectra
of He--Ar comparison arcs were obtained to calibrate the wavelength
scale. The spectrophotometric standard stars GD~108, GD~50, and
G~60-54 were observed with a 10\arcsec\ slit width at the beginning,
middle, and end of the night for flux calibration.

The reduction of all data was performed using the standard data
reduction packages MIDAS\footnote{ MIDAS is an acronym for the
European Southern Observatory package -- Munich Image Data Analysis
System.} and IRAF\footnote[2]{IRAF is distributed by National Optical
Astronomical Observatories, which is operated by the Association of
Universities for Research in Astronomy, Inc., under cooperative
agreement with the National Science Foundation}.  Cosmic ray hits were
removed from the 2D spectral frame in MIDAS.  Using IRAF tasks in the
{\em ccdred}\ package, we subtracted the bias and performed flat-field
corrections.  After that the 2D spectrum was wavelength-calibrated and
the night sky background was subtracted.  Using our data of the
spectrophotometry standard stars, the intensities of the 2D spectrum
were transformed to absolute fluxes.  One-dimensional spectra were
extracted such as to get total emission line fluxes.  All sensitivity
curves observed during the night were compared.  We found the final
curves to have a precision better than 2\% over the whole optical
range, except for the region blueward of 4000~\AA\ where the
sensitivity drops off rapidly.

All emission lines were measured applying the MIDAS programs described
in Kniazev et al.\, (\cite{K00},\cite{SHOC}):  These programs determine the
location of the continuum, perform a robust noise estimation, and fit
separate lines by a single Gaussian superimposed on the
continuum-subtracted spectrum.  Some overlapping lines were fitted
simultaneously as a blend of two or more Gaussian features: the
H$\alpha$  $\lambda$6563 and [\ion{N}{ii}] $\lambda\lambda$6548,6584
lines, the [\ion{S}{ii}] $\lambda\lambda$6716,6731 lines, and the
[\ion{O}{ii}] $\lambda\lambda$7320,7330 lines.  The emission lines
He~{\sc i} $\lambda$5876, [\ion{O}{i}] $\lambda$6300, [\ion{S}{iii}]
$\lambda$6312, [\ion{O}{i}] $\lambda$6364, H$\alpha$  $\lambda$6563,
[\ion{N}{ii}] $\lambda\lambda$6548,6584, \ion{He}{i} $\lambda$6678,
and [\ion{S}{ii}] $\lambda\lambda$6716,6731 were detected
independently in both grism spectra. Their intensities were averaged
after applying weights inversely proportional to the errors of the
measurements.  The quoted errors of single line intensities include
some components that are summed up in quadrature.  The total errors
were propagated in the calculations and are included in the
uncertainties of the element abundances and all derived parameters
presented here.

\begin{table}[hbtp]
\centering{
\caption{Line intensities in the observed spectra of the Fornax PN}
\label{t:Intens1}
\begin{tabular}{lcc} \hline
\rule{0pt}{10pt}
$\lambda_{0}$(\AA) Ion                  & F($\lambda$)/F(H$\beta$)&I($\lambda$)/I(H$\beta$) \\ \hline
3868\ [Ne\ {\sc iii}]\                    & 0.2428$\pm$0.0149 & 0.2434$\pm$0.0154 \\ 
3889\ He\ {\sc i}\ +\ H8\                 & 0.2028$\pm$0.0072 & 0.2269$\pm$0.0101 \\
3967\ [Ne\ {\sc iii}]\ +\ H7\             & 0.2441$\pm$0.0076 & 0.2697$\pm$0.0107 \\
4026\ He\ {\sc i}\                        & 0.0208$\pm$0.0019 & 0.0208$\pm$0.0019 \\ 
4101\ H$\delta$\                          & 0.2400$\pm$0.0072 & 0.2624$\pm$0.0098 \\
4267\ C\ {\sc ii}\                        & 0.0102$\pm$0.0012 & 0.0102$\pm$0.0012 \\
4340\ H$\gamma$\                          & 0.4464$\pm$0.0129 & 0.4627$\pm$0.0145 \\
4363\ [O\ {\sc iii}]\                     & 0.0371$\pm$0.0016 & 0.0369$\pm$0.0016 \\ 
4388\ He\ {\sc i}\                        & 0.0057$\pm$0.0009 & 0.0056$\pm$0.0009 \\ 
4471\ He\ {\sc i}\                        & 0.0488$\pm$0.0019 & 0.0484$\pm$0.0019 \\ 
4658\ [Fe\ {\sc iii}]\                    & 0.0032$\pm$0.0013 & 0.0032$\pm$0.0013 \\ 
4686\ He\ {\sc ii}\                       & 0.0028$\pm$0.0009 & 0.0028$\pm$0.0009 \\ 
4713\ [Ar\ {\sc iv]}\ +\ He\ {\sc i}\     & 0.0082$\pm$0.0010 & 0.0081$\pm$0.0010 \\
4740\ [Ar\ {\sc iv]}\                     & 0.0008$\pm$0.0004 & 0.0007$\pm$0.0004 \\ 
4861\ H$\beta$\                           & 1.0000$\pm$0.0286 & 1.0000$\pm$0.0291 \\
4922\ He\ {\sc i}\                        & 0.0141$\pm$0.0010 & 0.0139$\pm$0.0010 \\ 
4959\ [O\ {\sc iii}]\                     & 1.8359$\pm$0.0524 & 1.8101$\pm$0.0523 \\ 
5007\ [O\ {\sc iii}]\                     & 5.4626$\pm$0.1555 & 5.3824$\pm$0.1554 \\ 
5048\ He\ {\sc i}\                        & 0.0030$\pm$0.0012 & 0.0030$\pm$0.0012 \\ 
5538\ [Cl\ {\sc iii}]\                    & 0.0013$\pm$0.0007 & 0.0013$\pm$0.0007 \\
5876\ He\ {\sc i}\                        & 0.1464$\pm$0.0043 & 0.1428$\pm$0.0044 \\ 
6312\ [S\ {\sc iii}]\                     & 0.0054$\pm$0.0006 & 0.0052$\pm$0.0006 \\ 
6548\ [N\ {\sc ii}]\                      & 0.0194$\pm$0.0009 & 0.0188$\pm$0.0009 \\ 
6563\ H$\alpha$\                          & 2.9620$\pm$0.0852 & 2.8745$\pm$0.0909 \\
6584\ [N\ {\sc ii}]\                      & 0.0578$\pm$0.0034 & 0.0560$\pm$0.0034 \\ 
6678\ He\ {\sc i}\                        & 0.0408$\pm$0.0015 & 0.0395$\pm$0.0015 \\ 
6717\ [S\ {\sc ii}]\                      & 0.0044$\pm$0.0007 & 0.0043$\pm$0.0007 \\ 
6731\ [S\ {\sc ii}]\                      & 0.0047$\pm$0.0007 & 0.0045$\pm$0.0007 \\ 
7065\ He\ {\sc i}\                        & 0.0538$\pm$0.0019 & 0.0519$\pm$0.0020 \\ 
7136\ [Ar\ {\sc iii}]\                    & 0.0596$\pm$0.0021 & 0.0574$\pm$0.0022 \\ 
7236\ [Ar\ {\sc iv}]\                     & 0.0148$\pm$0.0012 & 0.0142$\pm$0.0012 \\ 
7281\ He\ {\sc i}\                        & 0.0083$\pm$0.0009 & 0.0080$\pm$0.0008 \\ 
7320\ [O\ {\sc ii}]\                      & 0.0089$\pm$0.0009 & 0.0086$\pm$0.0009 \\ 
7330\ [O\ {\sc ii}]\                      & 0.0085$\pm$0.0011 & 0.0082$\pm$0.0011 \\ 
7751\ [Ar\ {\sc iii}]\                    & 0.0145$\pm$0.0011 & 0.0140$\pm$0.0011 \\ 
8438\ P18\                                & 0.0022$\pm$0.0006 & 0.0021$\pm$0.0006 \\
8467\ P17\                                & 0.0027$\pm$0.0006 & 0.0025$\pm$0.0006 \\
8502\ P16\                                & 0.0039$\pm$0.0007 & 0.0038$\pm$0.0007 \\
8545\ P15\                                & 0.0080$\pm$0.0009 & 0.0076$\pm$0.0009 \\
8598\ P14\                                & 0.0058$\pm$0.0009 & 0.0056$\pm$0.0009 \\
8665\ P13\                                & 0.0080$\pm$0.0009 & 0.0077$\pm$0.0009 \\
  & & \\
C(H$\beta$)\ dex          & \MC {2}{c}{0.03$\pm$0.01}  \\
F(H$\beta$)$^a$\           & \MC {2}{c}{160$\pm$3}     \\
Rad. vel.\ \kms\           & \MC {2}{c}{72$\pm$14}    \\
\hline  \hline
\MC{3}{l}{$^a$ Observed flux in units of 10$^{-16}$ ergs\ s$^{-1}$cm$^{-2}$.}\\
\end{tabular}
 }
\end{table}

\section{Results}
\label{txt:res}

The relative intensities of all detected emission lines as well as the
derived C(H$\beta$), the measured flux in the H$\beta$ emission line,
and the measured heliocentric radial velocity are given in Table
\ref{t:Intens1}.  The final 1D spectra are shown in
Figure~\ref{fig:PN_spectr}.  The electron temperature $T_{\rm e}$, the
number density $N_{\rm e}$, and the ionic and total element abundances
for O, N, S, Ne, Cl, Ar, and He were calculated in the same manner as
detailed in Kniazev et al.\, (\cite{Sextans}).  They are presented in
Table~\ref{t:Chem}.  The [\ion{O}{ii}] $\lambda$3727,3729 doublet was
not observed, and O$^+$/H$^+$ was calculated using the intensities of
the [\ion{O}{ii}] $\lambda$7320,7330 lines.  The contribution to the
intensities of the [\ion{O}{ii}] $\lambda$7320,7330 lines due to
recombination was taken into account following the correction from Liu
et al.\, (\cite{Liu00}).  The electron temperatures for different ions
were calculated following Garnett (\cite{Gar92}) or Stasi\'nska
(\cite{Stas90}) in the manner described in e.g., 
Izotov \& Thuan (\cite{IT98}).  The 
ionization correction factors ICF(A) for
different elements were calculated using the equations from
Torres-Peimbert \& Peimbert (\cite{TPP77}), Garnett (\cite{Gar90}),
and Izotov \& Thuan (\cite{IT98}) and references therein.
The latter are derived for \ion{H}{ii} regions, but for weak \ion{He}{ii}
there are only minor differences with the PNe ICFs
of Kingsburgh and Barlow (\cite{KB94}).
Some Paschen lines are detected in the red part of the spectrum.
The \ion{C}{ii} $\lambda$4267 recombination line is also detected in our
spectrum. The ionic abundance of C was derived using the calculations
of Davey et al.\, (\cite{Dav00}) for the case B effective recombination
coefficients, which include both  radiative and dielectronic
recombination processes. Ionic abundances derived from optical recombination
lines (ORLs) depend only weakly on the adopted temperature and are
essentially independent of $N_{\rm e}$. A temperature of
$T_{\rm e} = 10500\pm500$~K was assumed throughout.
The ICF(A) for C was calculated using Kingsburgh and Barlow (\cite{KB94}).
The ionic and total element abundance for C are presented in Table~\ref{tbl:PN_C}.

After accounting for the Milky Way foreground extinction A$_{\rm B}$ =
0\fm08 -- 0\fm10 in the direction of the Fornax dSph
(Schlegel et al.\, \cite{Schlegel98}) we do not see any additional
internal extinction for the PN or in the Fornax dSph at this position.
Our observed H$\beta$ flux of the PN is about the same as was measured by
Danziger et al.\, (\cite{Danz78}), who found m$_{5007}$ = 18\fm91
(Jacoby \cite{J89}).  Our heliocentric radial velocity V$_{hel}$ =
72$\pm$14 is far from the value of V$_{hel}$ = 10$\pm$40 km s$^{-1}$
from Danziger et al.\, (\cite{Danz78}), but within the uncertainties
it is consistent with the Fornax optical velocity of 53$\pm$3 km
s$^{-1}$ (Mateo \cite{Mat98}; see also Walker et al. \cite{Walker06}
and Battaglia et al.\ \cite{Battaglia06}) and with the velocities of
the Fornax globular clusters (Mateo et al.\, \cite{Mat91} and references
therein).  Battaglia et al.\ (\cite{Battaglia06}) find a line-of-sight
$3\sigma$ velocity dispersion of $13.7\pm0.4$ km s$^{-1}$ for Fornax'
field stars.

Some of the physical parameters of the PN in Fornax can be estimated in the
same way it was done by Kniazev et al.\, (\cite{Sextans}) for the PNe in
Sextans~A and Sextans~B and by Kniazev et al.\, (\cite{Ket05b}) for the PN in
Leo~A.  With the \ion{He}{ii} $\lambda$4686 and H$\beta$ emission lines the
effective temperature ($T_{\rm eff}$) of the PN central star can be estimated
using the relation from Kaler \& Jacoby (\cite{KJ89}).  The weakness of the
line returns the minimum temperature of this relation, which must be
considered an upper limit. Also, the relation assumes optical thickness of the
nebula which may be in doubt. The (\ion{He}{ii} $\lambda$4686)/(\ion{He}{i}
$\lambda$5876) line ratios and (\ion{He}{ii} $\lambda$4686)/(\ion{He}{i}
$\lambda$4471) can be reproduced only with cool central stars ($T_{\rm eff}
\sim 50$--60\,kK) and optical depth $\tau_{13.6} < 10$, according to the
models of Gruenwald \&\ Viegas (\cite{GV00}).  We therefore adopt $T_{\rm
eff}=55\rm \, kK$.

The total luminosity ($L/{\rm L_{\odot}}$) can be derived from Zijlstra \&
Pottasch (\cite{ZP89}).  This assumes an optically thick nebula. For the
adopted $T_{\rm eff}$, we find $L = 1.0 \cdot 10^3 \,\rm L_\odot$, which is low
for a cool PN central star. This is for a black-body model: a stellar
atmosphere model may give a somewhat higher luminosity.

We have also derived the luminosity using the effective temperature and
stellar magnitude. The latter is obtained from the continuum emission in the
spectrum: at 5500\AA, we find $V=20.13\pm0.05$. The absolute magnitude is $M_V
= -0\fm60$. Using the non-LTE models of Rauch (\cite{Rauch03}), we find that
for $T_{\rm eff}=50\,$kK this corresponds to $L=6500\,\rm L_\odot$ and for
$T_{\rm eff}=60\,$kK, $L=11300\,\rm L_\odot$.  The difference with the
luminosity derived above suggests that the nebula is not optically thick in
all directions.

\begin{table}[h]
\centering{
\caption{Physical Parameters and O, N, Ne, S, Ar, Cl, Fe and He
Abundances of the PN in Fornax}
\label{t:Chem}
\begin{tabular}{lc} \hline
\rule{0pt}{10pt}
Parameter                            & Value                 \\ \hline
$T_{\rm e}$(OIII)(K)\                & 10,200$\pm$160 ~~     \\
$T_{\rm e}$(OII)(K)\                 & 11,050$\pm$150 ~~     \\
$T_{\rm e}$(ArIII)(K)\               & 10,650$\pm$130~~      \\
$T_{\rm e}$(SIII)(K)\                & 10,650$\pm$130~~      \\
$T_{\rm e}$(ClIII)(K)\               & 10,650$\pm$130~~      \\
$N_{\rm e}$(SII)(cm$^{-3}$)\         & 750$\pm$270~~         \\
& \\
O$^{+}$/H$^{+}$($\times$10$^5$)\     & 1.411$\pm$0.121~~     \\ 
O$^{++}$/H$^{+}$($\times$10$^5$)\    & 17.750$\pm$0.999~~    \\
O/H($\times$10$^5$)\                 & 19.160$\pm$1.006~~    \\
12+log(O/H)\                         & ~8.28$\pm$0.02~~      \\ 
& \\ 
N$^{+}$/H$^{+}$($\times$10$^7$)\     & 8.020$\pm$0.452~~     \\ 
ICF(N)\                              & 13.577                \\
N/H($\times$10$^5$)\                 & 1.09$\pm$0.06~~       \\
12+log(N/H)\                         & 7.04$\pm$0.02~~       \\
log(N/O)\                            & $-$1.25$\pm$0.03~~    \\
& \\ 
Ne$^{++}$/H$^{+}$($\times$10$^5$)\   & 2.198$\pm$0.189~~     \\ 
ICF(Ne)\                             & 1.079                 \\ 
Ne/H($\times$10$^5$)\                & 2.37$\pm$0.20~~       \\
12+log(Ne/H)\                        & 7.38$\pm$0.04~~       \\
log(Ne/O)\                           & $-$0.91$\pm$0.04~~    \\
& \\ 
S$^{+}$/H$^{+}$($\times$10$^7$)\     & 0.172$\pm$0.024~~     \\ 
S$^{++}$/H$^{+}$($\times$10$^7$)\    & 9.033$\pm$1.129~~     \\ 
ICF(S)\                              & 3.075                 \\ 
S/H($\times$10$^7$)\                 & 28.31$\pm$3.47~~      \\
12+log(S/H)\                         & 6.45$\pm$0.05~~       \\
log(S/O)\                            & $-$1.83$\pm$0.06~~    \\
& \\ 
Ar$^{++}$/H$^{+}$($\times$10$^7$)\   & 4.099$\pm$0.198~~     \\ 
Ar$^{+++}$/H$^{+}$($\times$10$^7$)\  & 0.315$\pm$0.161~~     \\ 
ICF(Ar)\                             & 1.006                 \\ 
Ar/H($\times$10$^7$)\                & 4.44$\pm$0.26~~       \\
12+log(Ar/H)\                        & 5.65$\pm$0.03~~       \\
log(Ar/O)\                           & $-$2.63$\pm$0.03~~    \\
& \\ 
Cl$^{++}$/H$^{+}$($\times$10$^7$)\   & 0.154$\pm$0.055~~     \\ 
ICF(Cl)\                             & 2.599                 \\ 
Cl/H($\times$10$^7$)\                & 0.40$\pm$0.14~~       \\
12+log(Cl/H)\                        & 4.60$\pm$0.15~~       \\
log(Cl/O)\                           & $-$3.68$\pm$0.16~~    \\
& \\
Fe$^{++}$/H$^{+}$($\times$10$^7$)\   & 1.412$\pm$0.582~~     \\
ICF(Fe)\                             & 16.971                \\
log(Fe/O)\                           & $-$1.90$\pm$0.18      \\
$[$O/Fe$]$\                          & 0.48$\pm$0.18~~       \\
$[$Fe/H$]$\                          & $-$1.13$\pm$0.18~~    \\
& \\
He$^{+}$/H$^{+}$\                    & 0.0922$\pm$0.0020~~   \\
He$^{++}$/H$^{+}$\                   & 0.0002$\pm$0.0001~~   \\
He/H\                                & 0.0924$\pm$0.0020~~   \\
12+log(He/H)\                        & 10.97$\pm$0.01~~      \\
\hline
\end{tabular}
 }
\end{table}

\clearpage
\begin{table}[h]
\centering{
\caption{C Abundance of the PN in Fornax}
\label{tbl:PN_C}
\begin{tabular}{lc} \hline
\rule{0pt}{10pt}
Parameter          & Value                  \\ \hline
C$^{++}$/H$^{+}$($\times$10$^4$)\    & 9.82$\pm$1.44~~       \\
ICF(C)\                              & 1.08                  \\
C/H($\times$10$^3$)\                 & 1.06$\pm$0.16~~       \\
12+log(C/H)\                         & 9.02$\pm$0.07~~       \\
log(C/O)\                            & 0.74$\pm$0.09~~    \\
\hline
\end{tabular}
 }
\end{table}

\begin{table}[h]
\centering{
\caption{Physical parameters of the Fornax PN and its progenitor}
\label{tbl:PN_par}
\begin{tabular}{lc} \hline
\rule{0pt}{10pt}
Parameter          & Value                  \\ \hline
T$_{eff}$ (K)      & 55,000                 \\
log10(T$_{eff}$)   & 4.74                   \\
L/L$_\odot$        & 7000                   \\
log(L/L$_\odot$)   & 3.85                   \\
M/M$_\odot$        & $\sim$1.2              \\
t$_{MS}$ (Gyr)     & $\sim$7.0             \\
		   &                        \\
m$_{5007}$         & 19\fm01$\pm$0\fm02     \\
A$_{5007}$         & 0\fm07                 \\
                   &                        \\
V$_\ast$ (mag)     & 20\fm13$\pm$0\fm05     \\
\hline
\end{tabular}
 }
\end{table}

We finally can derive limits on the luminosity from dynamical considerations.
The density derived above, together with the H$\beta$ flux indicates a
mass of 0.33\,M$_\odot$ and a radius of 0.16\,pc (assuming constant density).
(The radius of the nebula has however not been directly confirmed.)  For an
expansion velocity of 20\,km\,s$^{-1}$ (Richer \cite{RL06}), this implies an
age of the ejecta of roughly 8000\,yr. The models of stellar temperature
versus dynamical age (Gesicki et al. \cite{Gesicki06}) show that $T_{\rm
eff}=55\,\rm kK$ is reached after this time for models with a stellar mass of
0.58\,M$_\odot$.  The luminosity of such a star is $L=5000\,\rm L_\odot$.

All three derivations have uncertainties. Combining the last two, we
adopt $L=7000\,\rm L_\odot$.

After that the mass of the progenitor is obtained using the
theoretical tracks of Vassiliadis \& Wood (\cite{VW94}).  According to
Section~\ref{txt:disc} the metallicity is Z $\sim$1/4~Z$_{\odot}$ for the PN
in Fornax, where the solar value of the oxygen abundance is 12+$\log$(O/H) =
8.66 (Asplund et al.\, \cite{Asp04}).  Hence the tracks for Z=0.004 (1/5 of
Z$_{\odot}$) were chosen, which are the tracks closest to the above
metallicity.  The approximate age of the PN progenitor (t$_{MS}$) is derived
using the evolutionary lifetimes of various phases of stars from Vassiliadis
\& Wood (\cite{VW93}).  All these derived parameters are summarized in
Table~\ref{tbl:PN_par}.  In addition, m$_{5007}$ (Jacoby \cite{J89}) and
A$_{5007}$ (Cardelli et al.\, \cite{CCM89}) were calculated.

\begin{figure}
\centering
\includegraphics[width=8.8cm,angle=0,clip=]{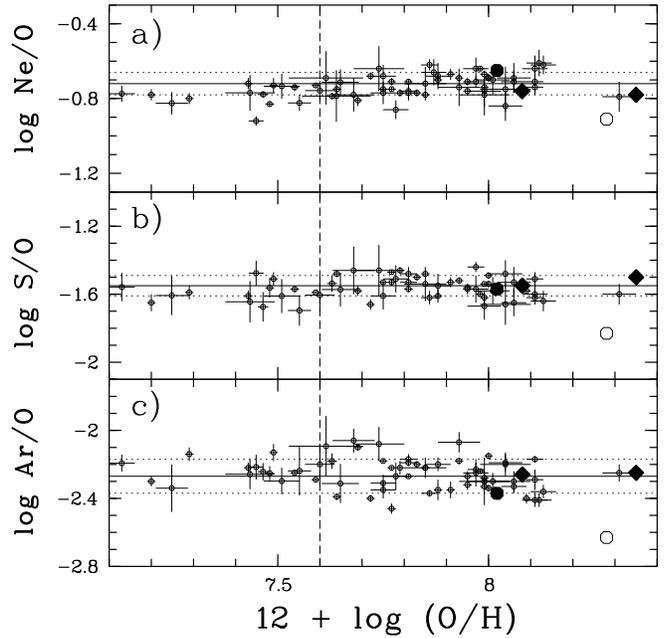}
   \caption{
$\alpha$-element-to-oxygen abundance ratios for log(Ne/O), log(S/O)
and log(Ar/O) with their 1$\sigma$ errors (short-dashed lines) from
Izotov \& Thuan (\cite{IT99}) as a function of oxygen abundance.  The
long-dashed line shows the value 12+log(O/H) = 7.60 above which these
abundance ratios were calculated.  Data for \ion{H}{ii} regions from
Izotov \& Thuan (\cite{IT99}) and from Kniazev et al.\,
(\cite{Kniazev03}) are overplotted.  Our calculated ratios for the PN
in Fornax are shown as open circles.  The same ratios corrected for
self-pollution in oxygen are shown as filled circles.  For comparison
data for the two of the PNe in the Sagittarius dSph 
galaxy (Dudziak et al.\, \cite{Dudziak00}) are shown as filled lozenges.
The chemical abundances for Sagittarius' PNe were recalculated with the
same method based on published spectral data.}
\label{fig:Abun_ratios} 
\end{figure}

\section{Discussion}
\label{txt:disc}

\subsection{Element Abundances of the PN in Fornax}

Danziger et al.\, (\cite{Danz78}) carried out the first spectroscopic
observations of the PN in the Fornax dSph galaxy.  They measured the
[\ion{O}{iii}] $\lambda$4363 line with an accuracy $\sim$30\% and
calculated 12+log(O/H) = 8.51 as well as He, Ar, and N abundances.
Their data permitted them only to present an upper limit for the
[\ion{S}{ii}] $\lambda\lambda$6717,6731 lines.  Maran et al.\,(\cite{Maran84})
recalculated the oxygen abundance (12+log(O/H) =
8.38) as well as the abundances of He, Ar, N, and S based on line
intensities from Danziger et al.\, (\cite{Danz78}).  Walsh et al.\,
(\cite{Walsh97}) tried also to observe the spectrum of the PN in
Fornax, but they did not detect the [\ion{O}{iii}] $\lambda$4363 line
nor the [\ion{S}{ii}] $\lambda\lambda$6717,6731 lines.  Richer \&
McCall (\cite{RM95}) estimated an oxygen abundance of the PN in Fornax
that is 0.4 dex lower than the one calculated by Maran et
al.\, (\cite{Maran84}).  These authors exploit this correction
by using the fact that the oxygen
abundance can be slightly higher for the brightest PN of a galaxy.
However, Walsh et al.\, (\cite{Walsh97}) argued that if a galaxy has
only one PN, then such a correction is not applicable.

Our data are considerably better (see Figure~\ref{fig:PN_spectr}) than
the earlier spectra, yielding a signal-to-noise ratio of 22 for the
[\ion{O}{iii}] $\lambda$4363 line.  With our new data we have for the
first time detected the weak [\ion{S}{ii}] $\lambda\lambda$6717,6731
lines (I(6717+6731) $\approx$ 0.01 I(H$\beta$)) and determined the electron
number density.  We obtain an oxygen abundance of 12+log(O/H) = 8.28
for the Fornax PN.

In \ion{H}{ii} regions, the oxygen that is seen has been produced by the same
massive stars that produced the alpha-process elements neon, sulfur and
argon. Therefore log(Ne/O), log(S/O), and log(Ar/O) should be constant and
show no dependence on the oxygen abundance.  These $\alpha$-element-to-oxygen
abundance ratios were measured by Izotov \& Thuan (\cite{IT99}) in \ion{H}{ii}
regions in blue compact galaxies and found to be log(Ne/O) = $-0.72\pm0.06$,
log(S/O) = $-1.55\pm0.06$ and log(Ar/O) = $-2.27\pm0.10$ for 12+log(O/H)$ >
7.60$ as shown in Figure~\ref{fig:Abun_ratios}.  For \ion{H}{ii} regions the
log(Cl/O) ratio also does not show any significant increase with increasing
oxygen abundance (Guseva et al.\, \cite{Gu03}; Esteban et al.\, \cite{Est98},
\cite{Est99}).

In contrast to \ion{H}{ii} regions some elemental abundances in PNe are
affected by the nucleosynthesis in the PN progenitors.  Newly synthesized
material can be dredged up by convection in the envelope, significantly
altering abundances of He, C, and N in the surface layers during the evolution
of the PN progenitor stars on the giant branch and asymptotic giant branch
(AGB).  But also a certain amount of O can be mixed in during the thermally
pulsing phase of AGB evolution.  For example, Kingsburgh \& Barlow
(\cite{KB94}) found that the O abundance is altered by $\sim$0.2 dex by AGB
evolution.  Kniazev et al.\, (\cite{Sextans}) studied Type~I PNe in Sextans~A
and found significant self-pollution of the PN progenitor by a factor of
$\sim$10 in oxygen. Pequignot et al.\ (\cite{P00}) show that the third
dredge-up increases the O-abundance at low metallicity but decreases it at
high metallicity, due to the specific abundance ratios in the dredge-up
material. Marigo (\cite{M2001}) also found this in theoretical dredge-up
calculations.

Our new measurements of O, Ne, Ar, and S abundances for the PN in Fornax are
shown in Figure~\ref{fig:Abun_ratios} as empty circles.  All log(Ne/O),
log(S/O), and log(Ar/O) ratios for this PN are systematically below the values
that were determined for \ion{H}{ii} regions.  This can be easily explained by
additional enrichment in oxygen in the progenitor of the Fornax PN.  Using
both the abundance ratios of Izotov \& Thuan (\cite{IT99}) and our observed
ratios this self-pollution can be calculated as weighted average $\delta$O =
0.27$\pm$0.10 dex.  After correction the resulting oxygen abundance
12+log(O/H) is 8.01 dex.  The corrected ratios (log(Ne/O)$_{\rm corr} =
-$0.64, log(S/O)$_{\rm corr} = -$1.56 and log(Ar/O)$_{\rm corr} = -$2.36) are
consistent with the values in \ion{H}{ii} regions, showing that a change in O
suffices. The corrected ratios are shown in Figure~\ref{fig:Abun_ratios} as
filled circles.  The chemical abundances for two of the PNe in the Sagittarius
dSph galaxy were recalculated with the same method on the base of published
spectral data.  Our Figure shows that the abundances of both PNe in
Sagittarius are located in the area where they are expected to lie.  We note
here that according to our calculations these PNe have different oxygen
abundances.  Dudziak et al.\, (\cite{Dudziak00}) find from detailed
photoionization modeling that their abundances are very similar.  The Sgr PN
Wray~16-423 has 12+log(O/H) = 8.35$\pm$0.02 according to our calculations, a
value that is very close to the previously published 8.33$\pm$0.02.  But the
Sgr PN He~2-436 has 12+log(O/H) = 8.08$\pm$0.02, which differs from the
earlier result of 8.36$\pm$0.06 (Dudziak et al \cite{Dudziak00}.

Our abundance for the PN in Fornax reflects the ISM metallicity at the epoch
of the PN progenitor formation about 8~Gyr ago (see Table~\ref{tbl:PN_par}).
Star formation histories for Fornax based on studies of its stellar
populations (e.g., Grebel \cite{Grebel99}; Tolstoy et al.\, \cite{Tol03})
indicate that the PN progenitor in Fornax originated at the time the star
formation rate in Fornax reached its peak.  Our [Fe/H] result is based on
the faint line [\ion{Fe}{iii}] $\lambda$4658 we derived as the result of WR
bump fitting; it may represent an upper limit. Using the fitted line strength
yields [Fe/H] = $-1.13\pm$0.18 for the Fornax PN. This is very close to the
photometrically derived mean metallicity of Fornax's old population $-$1.2 dex
(Grebel et al.\, \cite{GGH03}) and well within the range of metallicities
derived from spectroscopic measurements of red giant stars (Ca triplet) at
that epoch (see Fig.~23 in Battaglia et al.\ \cite{Battaglia06}), although
somewhat at the metal-rich part of that distribution.  The $\alpha$ elements
give a slightly higher abundance of [O/H]$\sim-0.7$.  At [Fe/H]$=-1$, chemical
evolution models give [O/Fe] for dwarf spheroidal galaxies in the range
0--+0.5 (Lanfranchi \&\ Matteucci \cite{LM2004}), declining with increasing
[Fe/H]: this is consistent with our value of [O/Fe]~$\approx0.5$.  We note
some caveats on this apparent agreement. First, the models do not fit the Sgr
dSph galaxy well, where the stars show a [O/Fe]~$\approx 0$ even at low
metallicity. Second, our value is based on an indirect line identification
which requires confirmation: if an upper limit, a lower Fe abundance is
derived. Third, Fe is strongly depleted in many PNe (e.g., Rodr\'iguez \&
Rubin \cite{RR05}), either due to dust condensation or due to iron processing
during the s-process nucleosynthesis. Although at face value we find no
evidence for depletion, it can not be ruled out either.

Our [Fe/H] or [O/H] value and PN progenitor age are consistent with the
(relatively flat) age-metallicity relation for the Fornax dSph presented by
Tolstoy et al.\, (\cite{Tol03}), by Pont et al.\, (\cite{Pont04}), and by
Battaglia et al.\ (\cite{Battaglia06}).  The PN predates the sudden steep rise
in metallicity seen for the youngest Fornax stars and for the youngest Sgr PN.
The young, metal-rich population present in Fornax and Sgr may show the effect
of accretion of enriched gas from the Milky Way (Zijlstra et al. \cite{Z06})
or from the proposed merger event (Coleman et al.\ \cite{Coleman04},
\cite{Coleman05}).

The dichotomy between the ORL and CEL (collisionally excited lines)
abundances is an open problem in nebular astrophysics (see Liu~\cite{Liu03},
and references therein).  Heavy-element abundances derived from ORLs are
systematically higher than those derived from CELs.  But for the Fornax PN,
the total C abundance 9.02$\pm$0.07 derived using the ORL $\lambda$4267 line
(see Table~\ref{tbl:PN_C}) is consistent within the quoted uncertainties with
the value 12+log(C/H) = 8.95 found by Maran et al.\,(\cite{Maran84}) on the
basis of measurements of the ultraviolet doublet line (CEL) \ion{C}{iii}]
$\lambda$1909. A higher S/N spectrum will be needed to identify any 
discrepancy between the ORL and CEL abundances in PN in Fornax.

\begin{figure} \centering
\includegraphics[width=6.3cm,angle=-90,clip=]{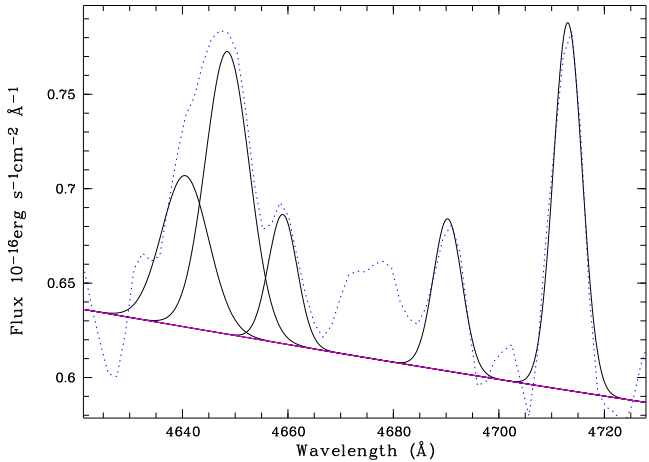} 
\caption{
Gaussian multicomponent fitting of the blue Wolf-Rayet bump.  The
observed spectrum in the fitted region and the continuum level are
shown.  The following lines from left to right are marked:
\ion{N}{iii} $\lambda$4640, \ion{C}{iii} $\lambda$4650,
[\ion{Fe}{iii}] $\lambda$4658, \ion{He}{ii} $\lambda$4686,
[\ion{Ar}{iv}]+\ion{He}{i} $\lambda$4713.  FWHMs for the
[\ion{Fe}{iii}] $\lambda$4658, \ion{He}{ii} $\lambda$4686 and
[\ion{Ar}{iv}]+\ion{He}{i} $\lambda$4713 lines are similar within the
uncertainties.  } 
\label{fig:WR_blue} 
\end{figure}

\subsection{The Central Star of PN in Fornax}

The Fornax PN shows a blue Wolf-Rayet bump in its spectrum: weak \ion{N}{iii}
$\lambda$4640 (EW = 1.3 \AA, FWHM = 10.4 \AA, flux = 0.7$\pm$0.3 10$^{-16}$
ergs\ s$^{-1}$cm$^{-2}$) and \ion{C}{iii} $\lambda$4650 (EW = 2.4 \AA, FWHM =
9.7 \AA, flux = 1.2$\pm$0.3 10$^{-16}$ ergs\ s$^{-1}$cm$^{-2}$), but no
detectable broad \ion{He}{ii} $\lambda$4686 and \ion{C}{iv} $\lambda$5810
emissions.
The detected lines are insufficient for a classification as [WC] star
and instead are consistent with the so-called {\it weak emission line stars}
(wels: Tylenda et al.\, \cite{TAS93}). The detection of \ion{N}{iii} supports
this, as this line is relatively common in {\it wels} but  rare for
[WC] stars. The relation between the {\it wels} and the [WC] stars
it not yet understood: Gesicki et al. (\cite{Gesicki06}) find the {\it
wels} are most common at a stellar temperature around 50\,kK, while the
[WC] stars are most common at lower and higher temperatures. This suggests
that the strength of the WR wind may be temperature (i.e. ionization)
dependent.The derived $T_{\rm eff}$ for the Fornax PN is consistent
with the {\it wels} peak of  Gesicki et al. (\cite{Gesicki06}).

It is worth noting that three of the PNe in the Sagittarius dSph galaxy
(Zijlstra et al. \cite{Z06}) also show Wolf-Rayet features in their spectra.
With the detected emissions the central star of the PN in Fornax is very
similar to the central star of the PN Wray 16-423 in Sagittarius that also was
classified as a {\it wels}.

The Wolf-Rayet spectral features in PNe spectra provide important
constraints on the evolutionary status of the central star.  As is
known, central stars of PNe evolve mostly either on hydrogen-burning
tracks or on helium-burning tracks, entering the white dwarf cooling
phase thereafter.  The hydrogen-burning phase is several times longer
than the helium-burning one (Vassiliadis \& Wood \cite{VW94})
and therefore the probability for the central star of the PN to be
detected in the helium-burning phase ranges from 20\% (Iben
\cite{Ib84}) to 25\% (Sch\"onberner \cite{Sc83}).  But since the
Wolf-Rayet central stars are known to be hydrogen-poor, with  high
probability such a star should be in the helium-burning phase.
This could be interpreted as that for low-metallicity stars, the final AGB
mass loss takes place preferentially during and immediately following a thermal
pulse. 

\section{Summary}
\label{txt:summ}

We present the highest signal-to-noise spectroscopy published to date of the
only known PN in Fornax.  The depth of our data permits us to measure line
ratios of elements not accessible in earlier studies.  We measured the
electron temperature, the electron density, and element abundances for He, N,
O, Ar, Ne, Cl, C, Fe and S, and derived the properties of the progenitor star.
These results are presented in several tables.  Except for nucleosynthetic
processing within the progenitor star, these data represent a snapshot of the
chemical composition in this dwarf galaxy approximately 8 Gyr ago when the
low-mass PN progenitor star was formed.  Roughly at this time the overall star
formation rate in Fornax reached its peak, giving rise to Fornax'
predominantly intermediate-age population.

We obtain an oxygen abundance of 12+log(O/H) = 8.28$\pm$0.02. According to our
analysis, this value should be corrected downward by 0.27+0.10 dex due to the
self-pollution of oxygen by the PNe progenitor. After this correction the
element abundance ratios Ne/O, S/O and Ar/O appear in a good overall accord
with the trends seen for H\,{\sc ii} regions in other galaxies.
The obtained
[Fe/H] value is slightly below the spectroscopically determined mean stellar
metallicity of Fornax, but this may be affected by gas-phase depletion of
iron. The abundances agree well with the stellar age-metallicity relation
in this galaxy.  The PN central star shows
Wolf-Rayet features in its spectrum, similar to the PNe in the Sgr dSph.

\begin{acknowledgements}

We thank the anonymous referee for comments which improved the
presentation of the manuscript. A.Y.K. thanks Albert Zijlstra for the
enormous support, help and useful discussions.
S.A.P. and A.G.P. acknowledge the partial support from Russian state
program ``Astronomy''.  E.K.G.\ thanks the Swiss National Science
Foundation for partial support through the grants 200020-113697 and
200020-105260.  This research has made use of the NASA/IPAC
Extragalactic Database (NED), which is operated by the Jet Propulsion
Laboratory, California Institute of Technology, under contract with
the National Aeronautics and Space Administration. We have also used
the Digitized Sky Survey, produced at the Space Telescope Science
Institute under government grant NAG W-2166.  This research has made
use of NASA's Astrophysics Data System Bibliographic Services. 

\end{acknowledgements}

\end{document}